\documentclass[12pt]{iopart}
\usepackage{graphicx}

\begin{document}

\title[High-frequency response of GaN]{High-frequency response of GaN in moderate electric and magnetic fields:
 Interplay between cyclotron and optical phonon transient time resonances}

\author{G I  Syngayivska, V V  Korotyeyev  and V A  Kochelap}

\address{Department of Theoretical Physics,
Institute of Semiconductor Physics,~Kiev 03028, Ukraine}

\ead{koroteev@ukr.net}

\begin{abstract}
We have studied the high-frequency properties of the non-equilibrium
electron gas in GaN samples subjected to electric and magnetic fields. Spectra
of the complex tensor of the dynamical mobility have been calculated for THz
frequency range. For the compensated GaN and low temperatures,
in the intervals of electric fields of the few $kV/cm$ and magnetic fields of the few $T$
the existence of the cyclotron and optical phonon transit-time resonances has
been identified. We have shown that interplay of two resonances gives rise to specific
spectra of THz transmission and absorption (or gain). We suggest that experimental
investigation of these effects will facilitate elaboration of field controlled devices
for THz optoelectronics.
\end{abstract}
accepted for publication in \SST

\maketitle

\section{Introduction}

Progress in the technology of group-III nitrides, discovery of unique material properties and
perspectives of high-power, and high-frequency applications~\cite{Pearton}
have inspired studies of high-field transport regimes
in these materials~\cite{Barker-1,Danil,Barker-2},
including the high-frequency phenomena~\cite{Zhang,Tsai,Guo}.
Most of these studies were focused to very high electric fields ($10...200~kV/cm$),
where the intervalley electron transfer and the Gunn effects are expected.
Also there is a necessity of understanding of hot electron kinetics
in the range of moderate electric fields ($1..10~kV/cm$), because of fundamental and
practical importance of such a case. Indeed, it is believed
(see recent review~\cite{review}) that in polar nitride materials
subjected to a moderately high  electric field, strong inelastic optical
phonon scattering can provide the so-called streaming transport regime.
The latter is characterized by considerable anisotropy of electron
distribution in the momentum space, quasi-saturation of the
current-field dependence, negative dynamic conductivity and  other
useful effects.

In general, the following conditions are favorable to realize the streaming
regime: (i) strong electron-optical phonon coupling, (ii) low temperatures which
provide optical phonon emission as dominant scattering mechanism of
hot electrons and  (iii) relatively low electron concentrations, when e-e scattering
is negligible. According to the analysis given in a number of the works~\cite{review},
these conditions can be met in the nitride materials, as well as in the nitride heterostructures~\cite{2D-1,2D-2}.
At the streaming regime, the electron motion is
nearly periodic: in an appropriate range of the {\it dc} electric field, an
electron accelerates quasiballistically until it reaches the optical phonon
energy, then the electron loses its energy by emitting an optical phonon
and starts the next period of acceleration\footnote {Note, this simplified picture can be traced back to papers by
Shockley ~\cite{Shockley} and Price  ~\cite{Price}.}.
Such a mechanism of the nearly periodic electron motion
leads to electric field induced resonances in high-frequency conductivity
[the so-called optical phonon transient time resonance (OPTTR)]~\cite{Andronov}.
The OPTTR provides an operating principle for electrically
pumped semiconductor sub-THz and THz sources~\cite{Andronov},
 that has been practically realized in n-InP crystals~\cite{THz-laser}.

Experimental observation of the OPTTR in the nitrides and its further device
implementation is impeded by the deficit of electron transport data for
the range of moderate electric fields, including high-frequency properties.

Additional useful information on properties of the nitrides can be
obtained by investigation of magnetic field effects on charge carriers,
for example, investigation of the electron cyclotron resonance (CR).
It is well known, that exploitation of the CR is a powerful
spectroscopic method that allows to determine important parameters
of the band structure and carrier dynamics (for a detail review see
ref.~\cite{CR-Otsuka}). Monitoring the cyclotron resonance
in an electric field, one can study intraband dynamic processes of nonequilibrium carriers.
For example, the cyclotron resonance technique was successfully used to study {\em hot
electrons and holes} in Ge~\cite{CR-Ge}, Si~\cite{CR-Si}, InSb~\cite{CR-InSb},
GaAs~\cite{CR-Otsuka} and other important semiconductor materials.
As the result, the detailed features of hot carrier kinetics have been understood.
These include: relaxation mechanisms and their parameters,
character of carrier distribution in the momentum space,
the dynamic conductivity, etc.
Recently, the cyclotron resonance was observed in the nitrides at
moderate and high magnetic fields (above 5 Tesla). Here the resonance
became apparent in THz-frequency range~\cite{CR-GaN-1,CR-GaN-2,CR-GaN-3,CR-GaN-4},
because of small sub-picosecond relaxation times characteristic for the nitrides.

In this paper by the use of the Monte-Carlo method, we study GaN bulk-like material in
moderate electric and magnetic fields. We focus on high-frequency response of GaN
under conditions, when both resonances, the CR and the OPTTR,  become apparent
simultaneously. We show that overlapping of the two resonances produces specific
spectra of THz transmission and reflection, as well as characteristic polarization effects.

\section{Model of electron transport}

We consider a bulk GaN sample of the cubic modification
with a given concentration of ionized impurity, $N_{i}$.
To exclude quenching effect on the OPTTR by electron-electron scattering
we suppose that the sample is compensated, thus the
electron concentration $n_{e}<N_{i}$. At electric fields of a moderate strength,
all electrons remain in the $\Gamma$ valley and can be characterized
by the parabolic dispersion law with the effective mass, $m^*$.
The $dc$ electric field, $\bf{F}$, and the magnetic field,
$\bf{B}$, are assumed to be parallel. Let the  both fields be  directed along
$OZ$ axis, ${\bf F}\parallel {\bf B}\parallel OZ$.
The magnetic field is treated as a classical one.
The $ac$ electric field is supposed to be harmonic,
${ \bf f}_{\omega}exp(-i\omega t)$,  with $\omega$ being the frequency.
To find the small-signal response, the $ac$ field is considered as a small
perturbation, $|{\bf f}_{\omega}| \ll |{\bf F}|$.

To calculate electron transport characteristics and, particularly,
frequency dependent dynamic mobility, $\mu_{\omega}$,
we use the single-particle algorithm of the  Monte-Carlo procedure~\cite{Reggiani,Zimmermann}.
For the simulation of the electron transport we take into account the main three
scattering mechanism: scattering by ionized impurities, acoustic phonons and polar-optical
phonons. For the  ionized impurity scattering, we exploit the "mixed" scattering model
that unifies both Brooks-Herring and Conwell-Weisskopf approaches.
This approach is more appropriate for the analysis of compensated
materials. The formulae of electron scattering probabilities
for these mechanisms can be found in ref.~\cite{Reggiani}.

In the limit of a classical  magnetic field the scattering probabilities
do not depend on this field.  However, the magnetic field determines the Lorentz force, so that
electron dynamics between sequential collisions is described by the equation :
\begin{equation}
m^* \dot{\bf v} = - e \left[ { \bf F}+{\bf f}_{\omega}exp{(-i\omega t})+
\frac{1}{c} [{\bf v} \times {\bf B}] \right],
\label{Eq}
\end{equation}
where $\bf{v}$ is the electron velocity, $e$ is the elementary charge,
$c$ is the light velocity. To avoid cumbersome expressions for ${\bf v} (t)$ during free flight between collisions
we present here only velocity modulation by the {\it ac} field, ${\bf f}_{\omega}$:
%\begin{equation}
\begin{eqnarray}
\textstyle
 \tilde{v}_{x}(t) = - \frac{ef_{\omega ,x}}{m^{*}}\frac{\omega\sin(\omega t_{0})}{\omega_{c}^{2}-\omega^{2}}  \cos(\omega_{c}\Delta t)-\frac{ef_{\omega ,x}}{m^{*}}\frac{\omega_{c}\cos(\omega t_{0})}{\omega_{c}^{2}-\omega^{2}} \sin(\omega_{c}\Delta t) +
\frac{ef_{\omega ,x}}{m^{*}}\frac{\omega\sin(\omega t)}{\omega_{c}^{2}-\omega^{2}},  \nonumber \\
%\scriptstyle
\textstyle
\tilde{v}_{y}(t)=-\frac{ef_{\omega ,x}}{m^{*}}\frac{\omega\sin(\omega t_{0})}{\omega_{c}^{2}-\omega^{2}}  \sin(\omega_{c}\Delta t)+\frac{ef_{\omega ,x}}{m^{*}}\frac{\omega_{c}\cos(\omega t_{0})}{\omega_{c}^{2}-\omega^{2}} \cos(\omega_{c}\Delta t)-
\frac{ef_{\omega ,x}}{m^{*}}\frac{\omega_{c}\cos(\omega t)}{\omega_{c}^{2}-\omega^{2}},  \nonumber \\
\textstyle
\tilde{v}_{z}(t)=\frac{ef_{\omega, z}}{m^{*}\omega}\left\{\sin(\omega t)-\sin(\omega t_{0})\right\},
\label{Eq1}
\end{eqnarray}
%\end{array}
%\label{Eq1}
%\end{equation}
where $t_0$ is an initial moment, $\Delta t= t-t_0$ and 
$\omega_{c}=e B/cm^{*}$ is the cyclotron frequency.
We used Eqs.~(\ref{Eq}) and (\ref{Eq1})  to apply the Monte-Carlo procedure to
the steady-state kinetics and high-frequency response of the electron gas
subjected to both electric and magnetic fields~\cite{Reggiani}.

First, we studied the steady-state distribution function of hot-electrons,
the drift velocity, $V_{d}(F)$, the average energy, $\overline{E} (F)$,
and other stationary electron characteristics in wide range of applied $dc$ electric fields,
when the magnetic field, $B$ is present. In Fig.\ref{fig0} we show an example of calculations of $V_{d}(F)$,
$\overline{E} (F)$ and distribution function. These results are discussed below in more details.

For the magnetic fields of a moderate
strength ($B < 10~T$), we found that these characteristics practically are
independent on the magnetic field.  This is a consequence of two main factors:
(i) for the parabolic electron dispersion and the parallel configuration of electric
and magnetic fields, electron motion along fields direction
and that in perpendicular plane are uncoupled; (ii) the magnetic field does not affect the energy balance of the electron system.
The detail analysis of the steady-state transport characteristics in compensated GaN
can be found in ref.~\cite{Sing2}.

In the presence of steady-state electric and magnetic fields, the
high-frequency response can be characterized by the dynamical mobility,
which is the tensor, ${\hat \mu}_{\omega}$. The dynamic mobility tensor defines the $ac$ electric current induced by
the $ac$ field, ${\bf J}_{\omega} = e n_e\hat{\mu}_{\omega} {\bf f}_{\omega}$. For the considered configuration of the
fields, this tensor has five non-zero components:
\begin{equation}
{\hat \mu}_{\omega}=\left( \begin{array}{ccc}
\mu_{\omega, xx}, &\mu_{\omega, xy} &0 \,\\
\mu_{\omega, yx}& \mu_{\omega, yy} &0 \\
0& 0& \mu_{\omega, zz}
\end{array} \right)\,,
\label{tensor}
\end{equation}
where only three components, $\mu_{\omega, xx}$,
$\mu_{\omega, xy}$ and $\mu_{\omega, zz}$,
are linearly independent. Other components are
$\mu_{\omega, yy}= \mu_{\omega, xx}$, $\mu_{\omega, xy}=- \mu_{\omega, yx}$.
Note, at $B =0$ nondiagonal components are equal to zero. In next Section
we present the analysis of spectra of all components of $\mu_{\omega}$ and
their  dependencies on the electric and magnetic fields.

\section{Calculation of the dynamic mobility}

Prior to discussion of results on the hot-electron dynamical mobility, $\hat{\mu}_{\omega}$,
we shall estimate ranges of the parameters favorable for observation of the OPTTR and the
CR effects in GaN.

\begin{figure}[h]
\centering
\includegraphics[height=6cm,width=8.5cm]{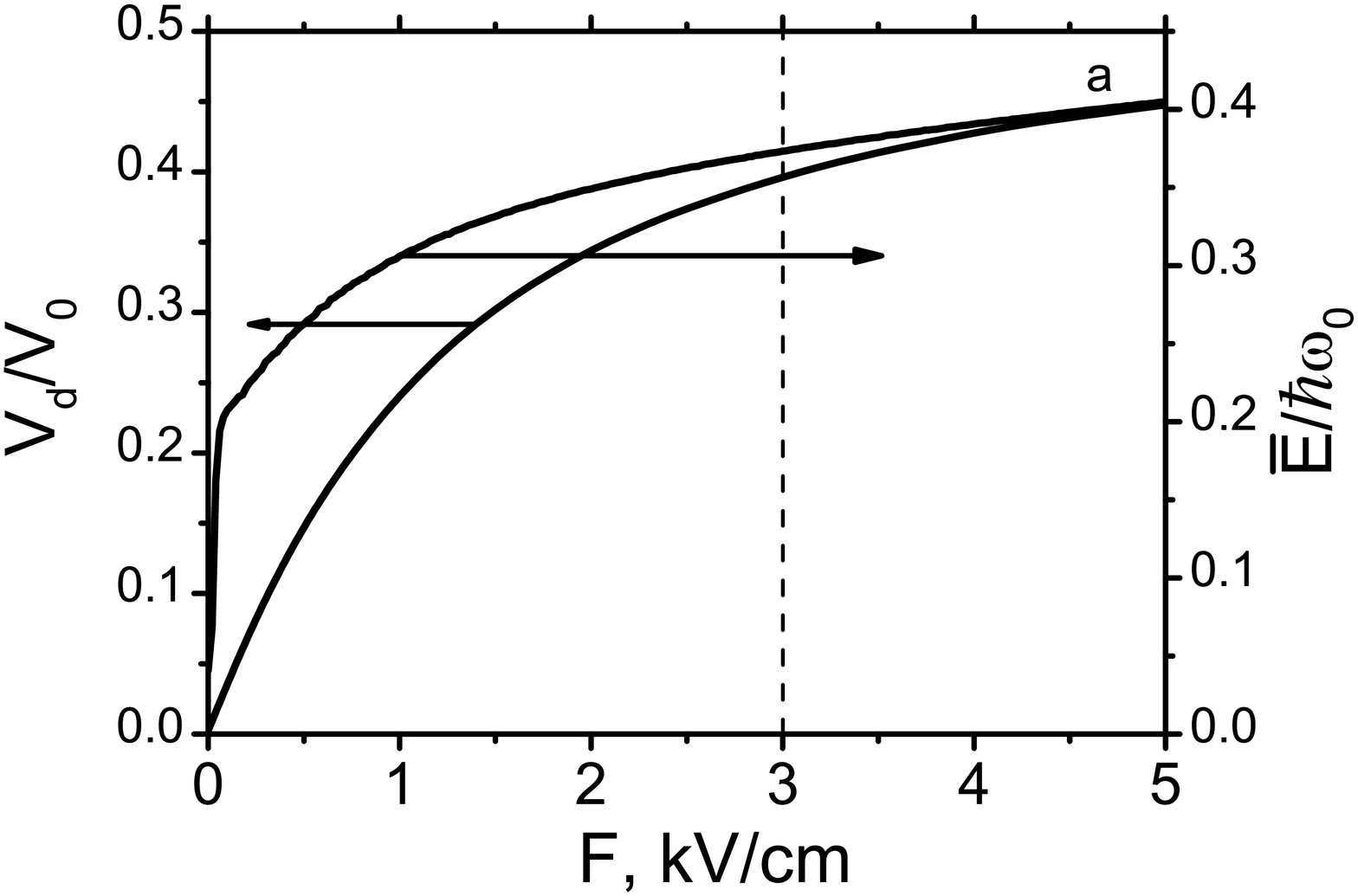}
\includegraphics[height=6.3cm,width=8cm]{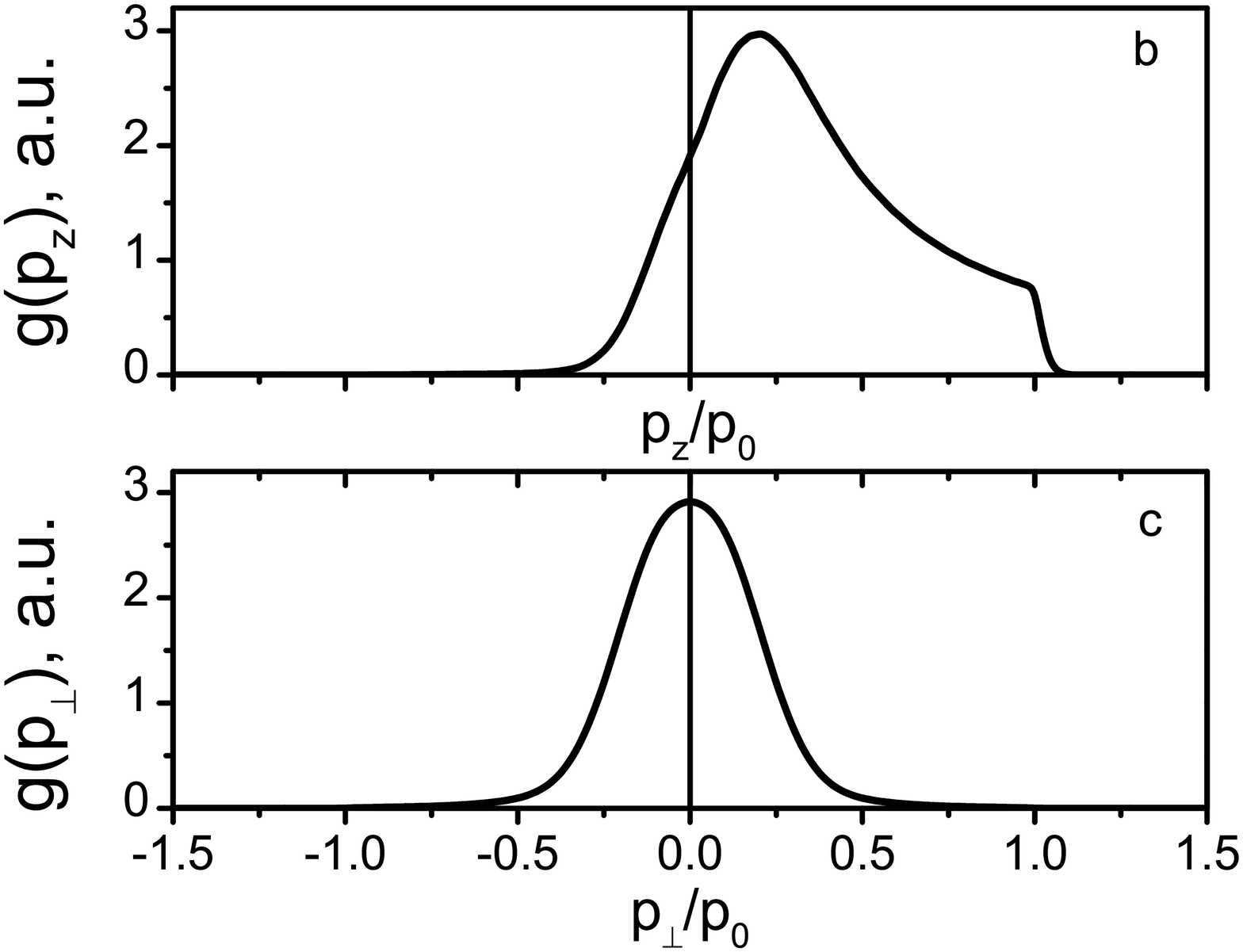}
\caption{ Panel (a) shows the field dependencies of the drift
velocity, $V_{d}(F)$, and averaged energy, ${\overline E}(F)$.
Panels (b) and (c) show the longitudinal  and transversal  profiles
of electron distribution function at $p_{\perp}=0$ and at $p_{z}=0$,
respectively, at $F=3\, kV/cm$. Characteristic velocity, $V_{0}=4\times 10^{7}\, cm/s$,
optical phonon energy $\hbar\omega_{0}=92 \, meV$, $p_0=\sqrt{2m^{*}\hbar\omega_{0}}$
is the characteristic momentum. }
\label{fig0}
\end{figure}

For concentrations of the electrons and ionized impurities we set
$n_{e}=10^{15}$ cm$^{-3}$ and $N_{i}=10^{16}$ cm$^{-3}$, respectively.
The lattice temperature is assumed to be 30 K.
For the GaN material parameters, such as effective mass, dielectric constants, deformation
potential, etc, we used the database from ref.~\cite{Shur}
 The low field mobility
was calculated to be $\approx 5000\,cm^2/V\,s$, which is well correlated with
the measured mobility in GaN samples of low dislocation density~\cite{Look}.
For these parameters, the streaming
transport regime and the OPTTR are the most pronounced in the range of the $dc$
electric fields $1...10\,kV/cm$. As seen from  Fig.~\ref{fig0}. (a),
in this range of applied $dc$ electric fields the drift velocity
and the averaged energy show the typical quasisaturation behavior. As the results of
streaming electron motion, the strongly asymmetric electron distribution function is
formed in momentum space. In Figs. \ref{fig0} (b) and (c), we present both
longitudinal and transversal profiles of the distribution function for the case
of developed streaming, $F=3\, kV/cm$. As seen, the transversal profile
of the distribution function remains symmetric. While for the longitudinal one
shows strong asymmetry: the electrons have momenta mainly directed along
the applied force. The streaming distribution
leads to the effect of OPTTR  in the frequency range $0.2-2\, THz$ ~\cite{review}.

Classical treatment of electron motion in the magnetic field
requires the condition $\hbar \omega_{c} \ll \overline{E}$.  For the above
range of the electric field, the average hot electron energy in GaN is estimated to be
${\overline E} < 30\,meV$. Then, the latter inequality gives
limitations: $\omega_{c} \ll 45$ THz and $B \ll 50$ T. In what follows, we
restricted ourself to the analysis of  moderate magnetic fields, $1...5\,T$,
which correspond to the cyclotron frequency $\omega_{c}/2\pi \approx 0.1...0.7\,THz$.

\subsection{General properties of ${\hat \mu}_{\omega}$ tensor}

\begin{figure}[h]
\centering
\includegraphics[height=9cm,width=9cm]{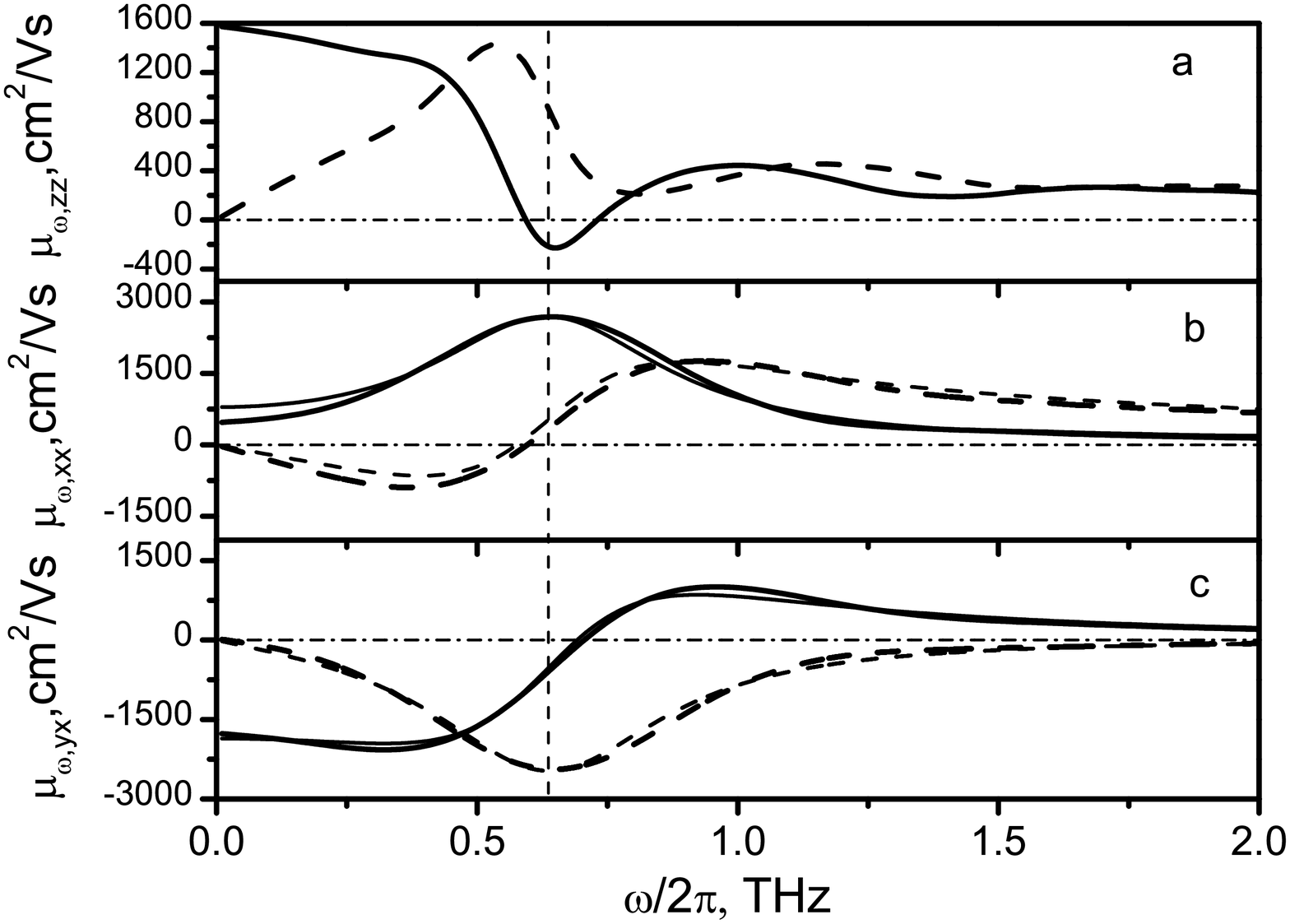}
\caption{(a), (b) and (c): Real (solid lines) and imaginary (dashed lines) parts of the components
$\mu_{\omega, zz}$,  $\mu_{\omega, xx}$ and $\mu_{\omega, yx}$
as functions of the frequency for  $F = 3\, kV/cm$ and  $B = 4.5\, T$.
Thin lines are fitting curves defined by Eqs.~\ref{Drude}.  }
\label{fig1}
\end{figure}

A representative example of  the dynamic mobility
of hot electrons under the developed streaming regime ($F = 3\,kV/cm$) is given in Fig.~\ref{fig1}.
The magnetic field is set to be $B = 4.5\,T$. In this figure, three independent tensor
components, $\mu_{\omega,xx},\,\mu_{\omega,zz},\,
\mu_{\omega,xy}$, are shown as functions of the frequency, $\omega$.

The 'longitudinal' (with respect to the applied fields) component, $\mu_{\omega,zz}$,
is practically independent on the magnetic field and demonstrates all signs
of the OPTTR effect: an oscillation behavior of both
$Re[\mu_{\omega,zz}]$ and $Im[\mu_{\omega,zz}]$, and a frequency "window",
where the real part of the dynamic mobility is negative, $Re[\mu_{\omega,zz}] < 0$.
Minimum of $Re[\mu_{\omega,zz}]$ reaches the value $ \approx - 200\, cm^2/Vs$ at $\omega/2 \pi = 0.63\,THz$.
 Notice, absolute values of $Re[\mu_{\omega,zz}]$ including the case $\omega \rightarrow 0$
 are much smaller than the low field mobility. This is a consequence of above mentioned
 quasi-saturation of the velocity - field characteristics.

The 'transverse' component, $\mu_{\omega,xx}$, and off-diagonal component, $\mu_{\omega,xy}$,
are dependent on both $F,\,B$ fields and demonstrate features proper to the CR. The
maximum of $Re[\mu_{\omega,xx}]$ and zero of $Im[\mu_{\omega,zz}]$ occur at
$\omega \rightarrow \omega_c$, $\omega_c/2 \pi \approx 0.63\,THz$  for the used parameters.
In the vicinity of $\omega_c$, we find that $Re[\mu_{\omega,xx}] $ is considerably larger than
$ |Re[\mu_{\omega,zz}]|$.  This is because all electrons contribute to the CR, while
exclusively streaming electrons contribute to the OPTTR effect.

We found that spectral dependencies of two tensor components, $\mu_{\omega,xx},\, \mu_{\omega,xy}$,
can be approximately described by the simple Drude-Lorentz model that takes into account
electron motion in a magnetic field. For this model, the components are
 \begin{eqnarray}
 \mu_{\omega, xx}^{(DL)}=\frac{e\tau^{*}}{m^*}\frac{1-i\omega\tau^{*}}
 {(1-i\omega\tau^{*})^{2}+(\omega_{c}\tau^{*})^{2}}\,, \nonumber\\
\mu_{\omega, yx}^{(DL)}=-\frac{e\tau^{*}}{m^*}\frac{\omega_{c}\tau^{*}}
{(1-i\omega\tau^{*})^{2}+(\omega_{c}\tau^{*})^{2}}\,,  \label{Drude}
\end{eqnarray}
where $\tau^*$ is an effective relaxation time. As seen from figures~\ref{fig1} (b) and  (c),
Eqs.~(\ref{Drude}) give a good fitting to the Monte Carlo result for $\tau^* = 0.6\,ps$.
Herewith,  the CR is well pronounced, because $\omega_c \tau^* \approx 2.4$.

\subsection{Electric-field dependence of the cyclotron resonance}

For the analyzed configuration of $F$ and $B$ fields, the electric field does not
affect the cyclotron motion of the electrons in the $\{x,\,y\}$-plane.
Instead, this field modifies the electron distribution and alters relaxation processes, and,
consequently, the effective relaxation time, $\tau^*$.

\begin{figure}[h]
\centering
\includegraphics[height=8cm,width=9cm]{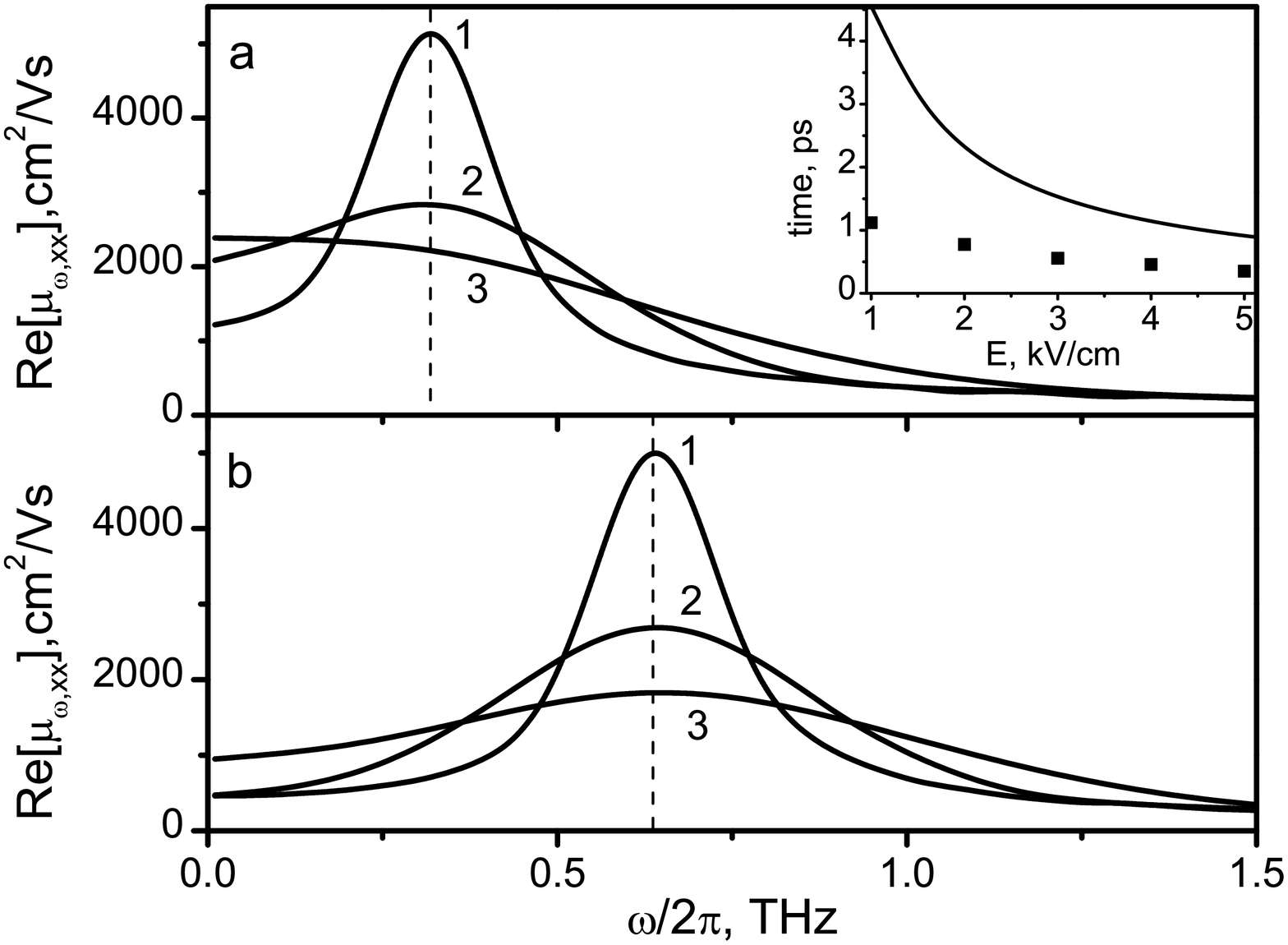}
\caption{ Real part of $\mu_{\omega, xx}$ as function of the frequency for
different $F$ and $B$ fields. (a): $B=2.2\, T$; (b): $B =4.5\, T$.
Curves 1, 2, 3 are calculated at $F=1, 3, 5\,\, kV/cm$, respectively. Vertical dashed lines
depict positions of cyclotron frequencies. In the insert we compare
$\tau^{*}(F)$  and $\tau_{tr}(F)$ dependencies (squires and solid line, respectively). }
\label{fig2}
\end{figure}

The results presented in Fig.~\ref{fig2} illustrate the electric field effect on the
cyclotron resonance.  One can see that, when the $F$ field varies, the frequency
of the resonance remains equal to $\omega_c$, while the sharpness of the resonance
degrades considerably with increase of the $F$ field magnitude.
For example, at $B = 2.2\,T$ ($\omega_c/2 \pi = 0.32\,THz$) the cyclotron peak in
$Re[\mu_{\omega, xx}]$ practically disappears when $F \geq 5\, kV/cm$ (see Fig.~\ref{fig2}(a)).
This is, obviously, due to an intensification of scattering processes. For the larger magnetic field,
$B =4.5\,T$, the CR remains observable for the range $F = 1...5\,kV/cm$, as shown in
Fig.~\ref{fig2}(b).   Therein we present also the $F$ field dependence of the effective
relaxation time, $\tau^*$, found by fitting Eqs.~\ref{Drude} to the Monte-Carlo results.
It is interesting, that for the $F$ field range, $1..5\,kV/cm$, when the streaming  regime
is formed, $\tau^*(F)$ dependence qualitatively correlates with the transit-time,
$\tau_{tr}$, during which a ballistically accelerated electron reaches the optical phonon energy,
$\hbar \omega_0$, i.e., $\tau_{tr} = \sqrt{2m^* \hbar \omega_0}/e F$. Thus under very fast process
of optical phonon emission, the width of the CR is determined rather by the dynamic characteristic,
$\tau_{tr}$.

\section{Coexistence and interplay of the CR and the OPTTR }

To measure microwave/optical effects in a magnetic field one uses, typically,
the Faraday and/or Voigt configurations of an incident wave and a magnetic
field. The CR can be observed for both configurations, while the OPTTR can be measured
only for the Voigt configuration (if ${\bf F}$ and ${\bf B}$ fields are parallel). To study coexistence
and interplay of both resonant effects  we will focus on the {\it Voigt configuration}.

For certainty, we set that  the incident electromagnetic wave propagates along
the OY axis and the high-frequency electric field, $\bf{{f}}_{\omega}$, lies in
the $\{ x,  z \}$-plane.  The incident wave is supposed to be linearly polarized.
Then, for the wave polarization along the $OX$ axis (${\bf f_{\omega}} || OX$) the CR
is observable, but the OPTTR is absent. To the contrary, for the wave polarization along
the $OZ$ axis (${\bf f_{\omega}} || OZ$) the CR is missed, but the OPTTR appears.
For other polarizations, both resonances coexist and affect each other.

\begin{figure}[h]
\centering
\includegraphics[height=8cm,width=9cm]{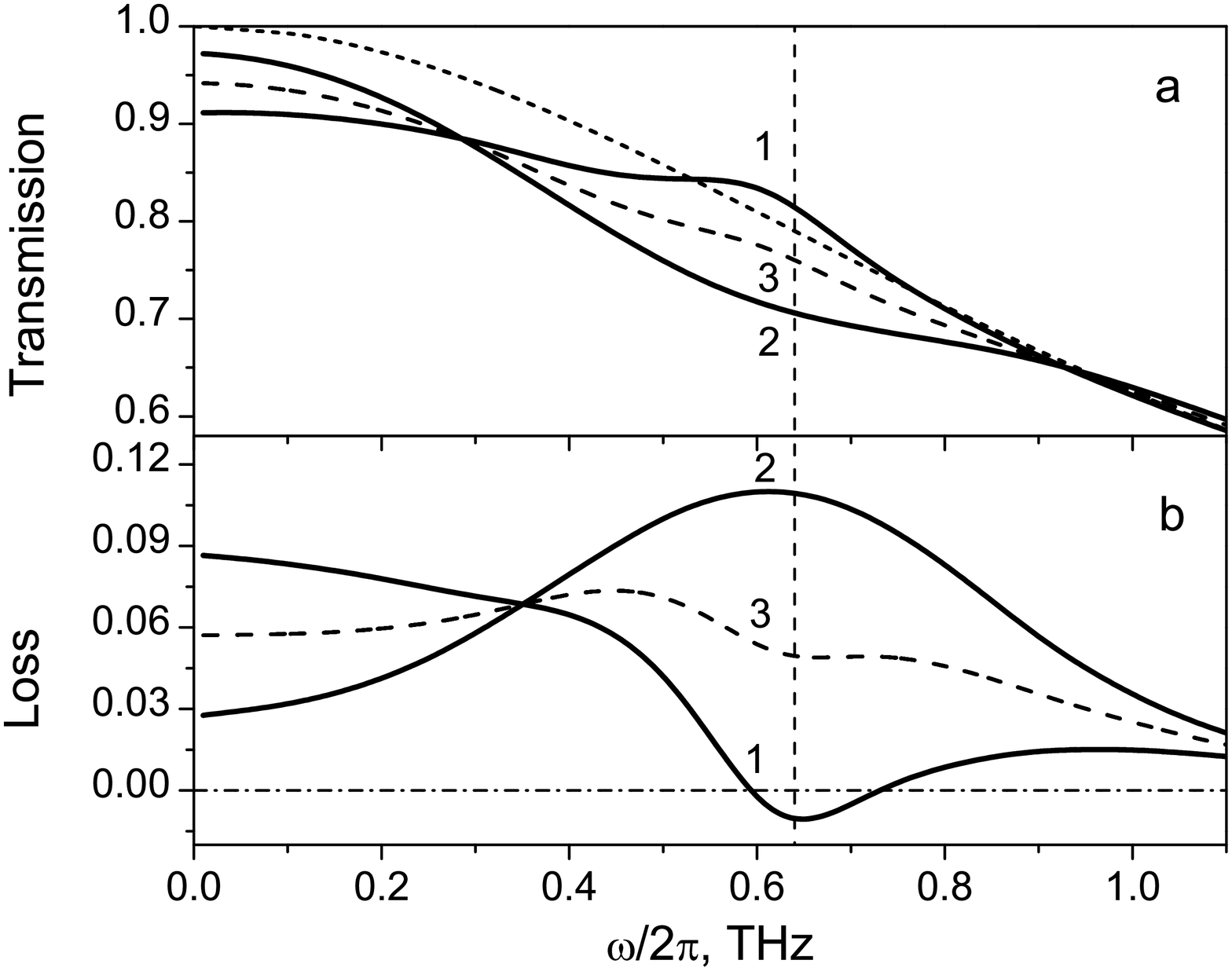}
\caption{ Transmission coefficient ( a) and loss coefficient (b) of a GaN plate.
Electron parameters and magnitudes of the $F$ and $B$ fields are the same as
in figure~\ref{fig1}. The plate thickness is $10\, \mu m$. Curve 1: ${\bf f} || OZ$,
curve 2: ${\bf f} || OX$, curve 3: the angle between ${\bf f}$ and the OZ axis is
$45^o$. Dotted line shows the lattice subsystem contribution.  }
\label{fig3}
\end{figure}

Interplay of the two resonances becomes apparent, for example, in transmission
spectra. These spectra comprise two contributions  from electron and lattice subsystems.
In additions, the spectra vary with the sample thickness because of reflection and interference
effects.  In Fig.~\ref{fig3}(a), we show the transmission spectra of a GaN sample in the
THz frequency range, where the CR and the OPTTR appear.  The electron parameters and
magnitudes of the $F$, $B$ fields  are the same as in Fig.~\ref{fig1}. The sample thickness
is set $10\,\mu m$, the dielectric permittivity is $8.9$.  For comparison, the contribution of the
lattice subsystem is shown separately. In accordance with the above said, the transmission
spectra depend on the polarization of the incident wave. For the polarization parallel to
the $F$, $B$ fields, the CR is absent, however due to the OPTTR
the electron contribution leads to an increase in the transmission coefficient (curve 1).
For the  perpendicular polarization,  the CR decreases the transmission (curve 3),
while there is no the OPTTR contribution.  The transmission spectra for an
intermediate polarization are illustrated by curve 3.

In Fig.~\ref{fig3}(b), we present frequency dependencies of the loss/gain coefficients
which append the data in Fig.~\ref{fig3}(a). The loss/gain coefficient defines
decrease/increase of the electromagnetic wave energy per unit time due to interaction with
the electrons in the GaN sample. One can see that  characteristic for the CR
decrease in the transmission coefficient is due to cyclotron absorption,
while sample 'bleaching' under OPTTR is relevant to a gain of the electromagnetic wave.

The loss/gain effects and their dependencies on the wave polarization can be
analyzed as follows. Let $\theta$ be an angle between the electric vector $\bf{f}_{\omega}$
and the $OZ$ axis. Then, the time-averaged $ac$ power density, that is received by
electrons from the electromagnetic wave, reads:
\begin{equation}
P_{\omega} (\theta) =\frac{1}{2}e n_{e} \left[\mu_{\omega,xx} sin^{2} (\theta)+
\mu_{\omega,zz}cos^{2}(\theta)\right]  |\bf{f}_{\omega}|^2\,.
\label{power}
\end{equation}
A negative value of $P_{\omega}$ indicates that an energy is transferred
from the electrons to the electromagnetic field and vice-versa, a positive $P_{\omega}$
means that the energy of the wave is absorbed by the electron subsystem.

\begin{figure}[h]
\centering
\includegraphics[height=8cm,width=9cm]{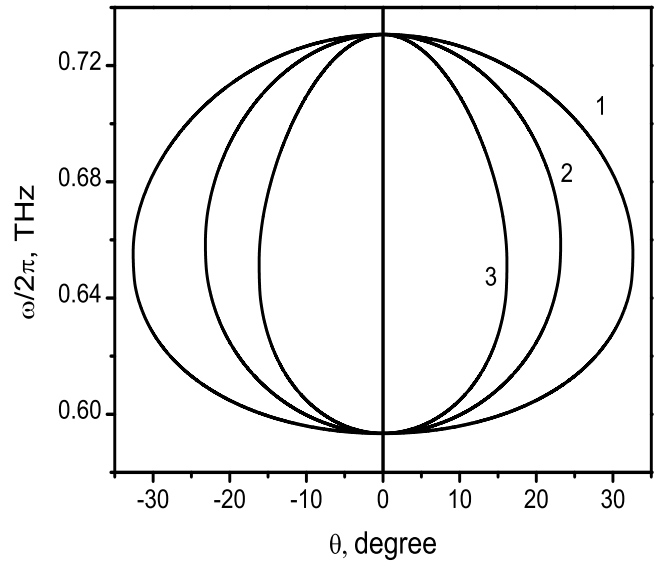}
\caption{Isolines $P_{\omega}(\theta)=0$ for different magnetic fields and $F= 3\,kV/cm$.
Curves 1, 2, 3 correspond to $H=0$, $2.2$, $4.5$ T, respectively. }
\label{fig4}
\end{figure}

In the frequency window, where $Re[\mu_{\omega,zz}] < 0$, the function $P_{\omega} (\theta)$
can be negative in some interval of the polarization angles $\theta$.
The condition $P_{\omega} (\theta) =0$ restricts the parameters $\theta\,,\omega$,
where the gain  due to the OPTTR occurs.
The actual space of these parameters can be illustrated in the $\{\theta,\,\omega \}$-plane, as
presented in Fig.~\ref{fig4} for different values of the $B$ field.  From this figure
it is seen that for $B=0$ the gain occurs  in the sufficiently wide angle sector,
$|\theta|<35^{\circ}$. With increasing $B$ this angle sector becomes narrow.
For example, at $H=4.5\, T$ the gain is possible only at $|\theta|<15^{\circ}$.
Thus, polarization dependencies of the gain effect are strongly affected by the CR.

\section{summary}

In this paper, we have investigated bulk-like GaN subjected to
electric and magnetic fields of moderate magnitudes, $1...5\,kV/cm$ and $1..5\,T$, respectively.
The parallel configuration of the electric and magnetic fields has been assumed.
With the use of the Monte Carlo method, we have calculated steady state and
high-frequency characteristics of the compensated GaN at low temperatures.
Particularly, we have focused on the situation, when two resonances,
the cyclotron resonance and the optical phonon
transient-time resonance,  become apparent simultaneously. We have calculated
spectral dependencies of all components of the tensor of dynamical mobility in
THz frequency range for different magnitudes of the electric and magnetic fields.

We have studied the CR for hot electrons, particularly, the electric field dependence of the CR
broadening.  For the analyzed range of the electric fields, where
the streaming  transport regime and the OPTTR effect are realized,
we have found that  the effective relaxation time, $\tau^*(F)$, determinative the broadening,
correlates with the transit-time, $\tau_{tr}$, during which a ballistically accelerated electron
reaches the energy of the optical phonon. That is, at very fast process of optical phonon
emission, the CR broadening is determined rather by the dynamic characteristic, $\tau_{tr}$.

We have determined spectral and field
dependencies of the OPTTR. Then, we have found that at electric fields of a
few $kV/cm$ and magnetic fields of a few $T$, well-developed cyclotron effect and
optical phonon transient-time effects may coexist in the frequency range of $0.5...1\,THz$:
they can be observed simultaneously for the Voigt configuration of the incident THz wave and
the fields. We have shown that interplay of two resonances gives rise to specific
spectra of THz transmission and absorption (or gain), as well as characteristic polarization effects.

It is important to make a few remarks on experimental observation
of the discussed effects in the THz frequency range. The spectral measurements in
the THz range can be performed by using recently developed {\em detection methods}
for ultrashort electric signals and {\em generation techniques} for sub-picosecond
transients of electric fields. The advanced ultrafast detection methods
mainly relay on two techniques: photoconductive sampling~\cite{PC-sampling-1} and electro-optic
sampling~\cite{EO-sampling-1,EO-sampling-2}. By the use of femtosecond laser pulses
with high repetition rates, these techniques allow for the
direct measurement of electric transients with sub-picosecond
time resolution. As for the generation techniques of sub-picosecond
electric field transients,  the most studied are THz pulses emitted by
transient photocurrents in a photoconductive switch and THz emission
generated by nonlinear optical processes (optical rectification,
four-wave mixing, etc). Additionally, time-resolved THz polarization and ellipsometry
methods are available. The techniques of generation and detection of
sub-picosecond electric field transients
were applied to GaN films to measure real and imaginary parts of the small signal
conductivity under equilibrium conditions~\cite{Zhang, Tsai, Guo}.
%~\cite{THz-GaN}
Apparently, the discussed THz techniques are generally compatible with
high-field measurement set-ups for magnetotransport in GaN films/plates.
Particularly, these techniques can be applied to study the CR and OPTTR for
the most interesting Voigt configuration of the fields.

Concluding, the presented results show that low temperature investigation of GaN
samples in moderate electric and magnetic fields can provide a useful information
on hot electron dynamic and relaxation, as well as reveal novel high frequency effects.
We suggest that experimental investigation of these effects would facilitate elaboration
of field controlled devices for THz optoelectronics, including THz emitters, amplifiers,
electro-optical modulators~\cite{modulators}, etc.

\ack
The authors acknowledge the support of bilateral cooperation
by the NASU/CNRS (Grant No. 24019) and Ukrainian
State Fund for Fundamental Researches (project No.F40.2/057).

\clearpage

\clearpage

\clearpage

\clearpage

\clearpage

\end{document}